\documentclass{aa}
\usepackage{graphicx}
\usepackage{txfonts}
\usepackage{subfigure}
\usepackage{multirow}
\usepackage{natbib}
\usepackage{array}
\usepackage{soul}
\usepackage[normalem]{ulem}

\usepackage{lscape}
\begin{document}
   \title{Unveiling the environment surrounding LMXB SAX~J1808.4$-$3658}


   \author{C. Pinto
          \inst{1}
          \and
          E.~Costantini\inst{2}
          \and
          A.~C.~Fabian \inst{1}
          \and
          J.~S.~Kaastra\inst{2,3}
          \and
          J.~J.~M. in 't Zand \inst{2}
          }

   \institute{Institute of Astronomy, Madingley Road, CB3 0HA Cambridge, 
             United Kingdom, \email{cpinto@ast.cam.ac.uk}.
         \and
             SRON Netherlands Institute for Space Research,
              Sorbonnelaan 2, 3584 CA Utrecht, The Netherlands.
         \and
             Astronomical Institute, Utrecht University,
             P.O. Box 80000, 3508 TA Utrecht, The Netherlands
         }
   \date{Received December 12, 2013 / Accepted January 30, 2014}

 
  \abstract
   {Low-mass X-ray binaries (LMXBs) are a natural workbench to study accretion disk phenomena and optimal background sources to measure elemental abundances in the Interstellar medium (ISM). In high-resolution XMM-\textit{Newton} spectra, the LMXB SAX~J1808.4$-$3658 showed in the past a neon column density significantly higher than expected given its small distance, presumably due to additional absorption from a neon-rich circumstellar medium (CSM).}
   {It is possible to detect intrinsic absorption from the CSM by evidence of Keplerian motions or outflows. For this purpose, we use a recent, deep (100\,ks long), high-resolution \textit{Chandra}/LETGS spectrum of SAX~J1808.4$-$3658 in combination with archival data.}
   {We estimated the column densities of the different absorbers through the study of their absorption lines. We used both empirical and physical models involving photo- and collisional-ionization in order to determine the nature of the absorbers.}
   {The abundances of the cold interstellar gas match the solar values as expected given the proximity of the X-ray source. For the first time in this source, we detected neon and oxygen blueshifted absorption lines that can be well modeled with outflowing photoionized gas. The wind is neon rich (Ne/O\,$\gtrsim$\,3) and may originate from processed, ionized gas near the accretion disk or its corona. The kinematics ($v=500-1000$\,km\,s$^{-1}$) are indeed similar to those seen in other accretion disks. We also discovered a system of emission lines with very high Doppler velocities ($v\sim24\,000$\,km\,s$^{-1}$) originating presumably closer to the compact object. Additional observations and UV coverage are needed to accurately determine the wind abundances and its ionization structure.}
   {}


   \maketitle
%

\section{Introduction}
\label{sec:introduction}

Accretion disks provide an important workbench to probe magnetized plasma dynamics, photoionization, thermal and ionization equilibria. The response of the ionized plasma to changes in the continuum and outflow kinematics may constrain the source that drives winds, the connection between jets, disk, and corona, and possibly the disk geometry. Low-mass X-ray binaries (LMXBs) consist of a neutron star (NS) or black hole (BH) in orbit with a $<$1 M$_{\odot}$ companion. These sources are characterized by X-ray emission due to hot gas accreted by the compact objects in a form of disks. Some of the NS binaries exhibit X-ray bursts when enough of this material accumulates on the surface of the NS, eventually giving rise to a runaway thermonuclear explosion. Bursts can last up to an hour, but they provide only a few\,\% of the total fluence (see e.g. \citealt{Lewin1993}). 

LMXBs display high flux and simple spectra, whose continuum is typically
well described with blackbody (BB) and powerlaw (PL) emission. The BB component arises from the thermal emission of the accretion disk around the neutron star. The powerlaw describes the energy gain of the disk soft photons by scattering in the accretion disk corona, a process known as Comptonization. A few sharp features are seen in the high-energy part of the spectra, e.g. above 1\,keV. Prominent Fe\,K emission lines are commonly seen in LMXB high-quality spectra and a clear asymmetry is seen in the line profile, as would be expected if the lines originate from the innermost region of the accretion disk and therefore are subject to strong relativistic effects (see e.g. \citealt{Cackett2012} and references therein). 

Most LMXB spectra do not exhibit intrinsic sharp absorption features in the soft X-ray energy band, but they are rich with absorption lines originating in the interstellar medium (ISM) that lies along their line-of-sight (LOS). The ISM influences the Galactic evolution through the exchange of matter with the stars and shows a complex structure consisting of phases at different equilibrium temperatures \citep[for a review, see][]{Draine2011}. The K-shell transitions of carbon, nitrogen, oxygen, neon, and magnesium, and the L-shell transitions of iron fall in the soft X-ray energy band. The launch of the {XMM-\textit{Newton}} and \textit{Chandra} satellites, provided with high spectral resolution gratings, permitted to determine interstellar ionization states and amounts of dust in the LOS towards several X-ray sources (see e.g. \citealt{paerels, devries, JuettI, costantini2005, Costantini2012, costantini, kaastra09, Lee2009, Pinto2010}).

Recently, \citet{Pinto2013}, hereinafter P13, have performed an extended analysis of the ISM towards nine LMXBs with high-quality {XMM-\textit{Newton}} spectra, extracted total abundances, dust depletion factors, and attempted to constrain dust chemistry. They also measured the abundance trends in the Galactic disk and confirmed the well known metallicity gradient (see e.g. \citealt{GradPedicelli}). 
However, their interstellar abundances significantly deviate from a monotonic distribution. The extreme case is given by the LMXB \object{SAX~J1808.4$-$3658} (hereinafter SAX~J1808) where the neon abundance exceeds by 5$\sigma$ the value predicted by the source location. P13 argue that this large scatter may be due to absorption by metal-rich material surrounding the X-ray sources. 

Absorption occurring near the source may be distinguished from Galactic absorption when velocity shifts or Doppler broadening are observed. However, the XMM-\textit{Newton} spectral resolution may not be high enough to resolve them. Recent work by \citet{Schulz2010} on the high-resolution \textit{Chandra} spectra of LMXB 4U\,0614$+$091 showed a variable Ne\,K edge with an average velocity smear of $\sim3500$\,km\,s$^{-1}$ implying a characteristic radius $<10^7$\,m, consistent with an ultra-compact binary (UCB) nature. The variability proves that the excess is intrinsic to the source, but it is not yet clear whether this is due to either neon~/~oxygen absorption or some other 
process. \citet{Costantini2012} found an absorption-line system in the RGS spectrum of the LMXB 4U1820--30, which may also be described as a $\sim1200$\,km\,s$^{-1}$ outflow. \citet{Ioannou2003} found evidence of a $\sim1500$\,km\,s$^{-1}$ \ion{C}{iv} outflow in the \textit{Hubble Space Telescope} (HST) UV spectrum of the LMXB X2127+119.

The X-ray transient SAX~J1808 is the first discovered accretion-powered millisecond X-ray pulsar in a LMXB (\citealt{zand1998, Wijnands1998}). It is relatively nearby ($d=3.5\pm0.1$\,kpc, see \citealt{Galloway2006}) and has a column density of interstellar medium $N_{\rm H} = (1.40\pm0.03)\,\times 10^{25}$\,m$^{-2}$ (see P13). This $N_{\rm H}$ value combined with the high source flux produces several strong interstellar absorption features in soft X-ray spectra. The source exhibits X-ray bursts showing photospheric expansion (see e.g. \citealt{Galloway2008}). These most likely originate in a flash of a pure helium layer that is produced by stable hydrogen burning. This source is thus particularly suitable to study X-ray bursts in great detail and in a large bandpass.  Therefore, \citet{zand2013}, hereinafter ZA13, observed SAX~J1808 with Chandra and RXTE during the November 2011 outburst. They detected a single thermonuclear (type-I) burst, the brightest yet observed by Chandra from any source, and the 
second-brightest observed 
by RXTE. 
However, they found no evidence for discrete spectral features during the burst $-$ though absorption edges have been predicted to be present in such bursts $-$ and argued that a greater degree of photospheric expansion may be required. The persistent spectrum is instead optimal to study the absorption edges produced by diffuse (ISM) and circumstellar (CSM) media.

In this work we used the deep observation of SAX~J1808 presented in ZA13, which was taken with the Advanced CCD Imaging Spectrometer (ACIS) in combination with the Low-Energy Transmission Grating (LETG) on board of {\textit{Chandra}} during a high-flux source state as well as the archival observations taken with the {\textit{Chandra}} Medium-Energy Grating (MEG) and the {XMM-\textit{Newton}} Reflection Grating Spectrometer (RGS).

The paper is organized as follows. In Sect.~\ref{sec:data} we present the data reduction. In Sect.~\ref{sec:spectra} we describe the spectral features and the fitting procedure. Alternative methods of analysis are treated in detail in Sect.~\ref{sec:spectral_models1} and \ref{sec:spectral_models2}. In Sec.~\ref{sec:variability} we analyze the spectral variability. All results and their comparison with previous work are discussed in Sect.~\ref{sec:discussion}. Conclusions are given in Sect.~\ref{sec:conclusion}.

\section{The data}
\label{sec:data}

In this work we use the medium-to-high quality gratings observations available for this source as given in Table~\ref{table:1}. The observations concern three different epochs between which SAX\,J1808 may have experienced significant spectral changes. The first {\textit{Chandra}} observation (epoch~1) has a short exposure and it is mainly used for comparative purposes. Most of our work focuses on the {XMM-\textit{Newton}} (epoch~2) and especially on the second {\textit{Chandra}} (epoch~3) observations.

\begin{figure}
  \begin{center}
      \subfigure{ 
      \includegraphics[bb=65 115 510 755, width=6cm, angle=-90]{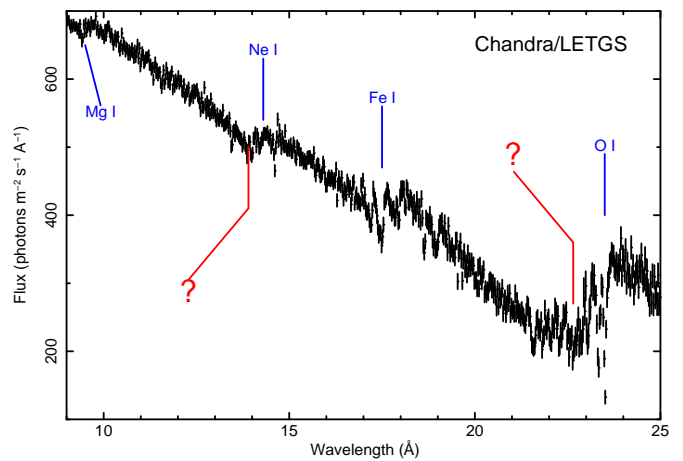}}
      \caption{\textit{Chandra}/LETGS spectrum. Absorption lines
      produced by neutral ISM gas are labeled. Two absorption features presumably originating near the source 
      are shown by a red question mark at 14.0 and 22.6\,{\AA}.}
          \label{fig:LETGS}
  \end{center}
\end{figure}

We reduced the {\textit{Chandra}} data following the standard threads provided by the {\textit{Chandra}} Interactive Analysis of Observations (CIAO) version 4.5. The bursting period was about 600\,s (see ZA13) and has been removed from the event file in order to obtain the spectrum of the persistent emission. The LETGS -1/+1 orders were combined with the CIAO task \textsl{add\_grating\_orders}. We show the LETGS spectrum of the persistent emission in Fig.~\ref{fig:LETGS}. The most relevant 
spectral features presumably produced by the ISM lying along the LOS are labeled. 
In order to check whether the bright X-ray burst altered the spectrum during the LETGS observation,
we divided the event file in two parts: {pre-burst (67.7\,ks) and post-burst (32\,ks, which includes the burst).}

The {XMM-\textit{Newton}} dataset was reduced with the Science Analysis System (SAS) version 13.0.1. We did not stack the RGS\,1 and 2 spectra in order to avoid spurious effects due to possible defects in their cross-calibration. However, the RGS spectra as obtained with the newer calibration files do not significantly differ from those shown in P13. We used the full 7--35\,{\AA} RGS spectra. We focus on the 7--27\,{\AA} MEG--LETGS wavelengths ranges because the \textit{Chandra} spectra are dominated by noise above 27\,{\AA}. We model the spectra with SPEX\footnote{www.sron.nl/spex} version 2.03.03 \citep{kaastraspex}. We scaled the interstellar abundances to the recommended protosolar or Solar System values of \citet{Lodders09}, to which we simply referred as \textit{solar abundances} throughout the paper. We adopted $\chi^2$-statistics and $1\,\sigma$ errors throughout the paper unless otherwise stated.

\begin{table}
\caption{Log of the observations used in this paper.}  
\label{table:1}    
\renewcommand{\arraystretch}{1.3}
\begin{center}
 \small\addtolength{\tabcolsep}{-2pt}
\scalebox{1}{%
\begin{tabular}{c c c c}     
\hline            
\hline            
                  Detector   &  ID $^{(a)}$ & Exposure (ks) $^{(b)}$ & Date \\  
\hline   
                 \textit{Chandra}/MEG         & 6297 & 14.4  & 2005-06-12\\
                 XMM-\textit{Newton}/RGS & 0560180601 & 63.7  & 2008-09-30\\
                 \textit{Chandra}/LETGS          & 13718 & 99.7  & 2011-11-08\\
\hline                
\end{tabular}}
\end{center}
$^{(a, \, b)}$ Identification number and net duration of the observation.\\
\end{table}

\section{Spectral features and fitting procedure}
\label{sec:spectra}

\subsection{Spectral features}

In Fig.~\ref{fig:LETGS} we show the neutral absorption edges of Mg ($\lambda=$ 9.5\,{\AA}), Ne ($\lambda=14.3$\,{\AA}), Fe ($\lambda=$ 17.15 and 17.5\,{\AA}), and O ($\lambda=$ 23\,{\AA}) of SAX~J1808 as seen with the LETGS. Several additional absorption lines are produced by mildly and highly ionized gas species like \ion{Ne}{ii-iii}, \ion{Ne}{ix}, and all the \ion{O}{ii-viii} series. The outer spectral regions do not show additional strong features. The \ion{N}{i} 1s--2p edge at 31.3\,{\AA} is weak because the background dominates above 28\,{\AA}.

Since dust may contain a significant amount of neutral iron and oxygen, it is expected to affect the spectral curvature between $17.1-17.6$\,{\AA} and $22.7-23.2$\,{\AA}. P13 indeed detected dust, measured depletion factors, and suggested possible compounds using the XMM-\textit{Newton}/RGS spectrum.

We also detect two interesting features at 14.0 and 22.6\,{\AA} which are difficult to ascribe to interstellar absorption. The 14.0\,{\AA} feature can be described as a blueshifted neon absorption line, while that at 22.6\,{\AA} is likely produced by blueshifted oxygen. The \ion{Ne}{ii} absorption line has a blueshifted wing with $v\sim500$\,km\,s$^{-1}$ (see also Fig.~\ref{fig:neon}).

\subsection{The fitting procedure}
\label{sec:procedure}

We modeled the spectra with two different approaches improving the empirical ISM models given in P13 and the physical models adopted by \citet{Pinto2012}, hereinafter P12. We first performed an empirical edge-by-edge analysis of the LETGS spectrum searching for any evidence of ionized gas and dust as well as of blue/red-shifted features. In fitting each edge the choice of the continuum is not critical and a single powerlaw with slope of about 2.2 is enough to reproduce it. We use the wavelength ranges 12.7--14.7\,{\AA}, 16--18.8\,{\AA}, and 18--26\,{\AA} to model the neon, iron, and oxygen edges, respectively.

Afterwards, we jointly fit the RGS and LETGS spectra by adopting a continuum consisting of blackbody plus powerlaw emission and a combination of photo- and collisional-ionization models for the absorbers. In all our fits we fix the hydrogen column density to $1.4\times10^{25}$m$^{-2}$ (which was measured in P13).

\section{Edge-by-edge spectral modeling with LETGS}
\label{sec:spectral_models1}

We first apply an \textit{empirical} model to measure the column densities without adopting a certain ionization balance. X-ray absorption lines in our spectra are too narrow to be resolved, which means that velocity widths and column densities are degenerate. We therefore fix the velocity dispersion of each absorber to a reasonable value (see below).

\subsection{Model description}

The atomic species that produce absorption lines in the spectrum have been modeled with the {\sl slab} model in SPEX, which gives the transmission through a layer of gas with arbitrary ionic column densities. A {\sl slab} component is required for the cold and warm interstellar gas which have a similar velocity dispersion of 10\,km\,s$^{-1}$ (see e.g. P12). Another {\sl slab} component accounts for the highly-ionized gas which may have higher turbulence, but its velocity width is forced to range between 10 and 100\,km\,s$^{-1}$.

The atomic species of the cold gas that produces continuum absorption or lines outside our wavelength range were described with the {\sl hot} model in SPEX, which calculates the transmission of a collisionally-ionized equilibrium plasma. For a given temperature and set of abundances, the model calculates the ionization balance and then determines all the ionic column densities by scaling to the prescribed total hydrogen column density. We freeze the temperature to the minimum value in SPEX ($0.5$\,eV) in order to mimic the neutral interstellar gas, but the 
abundances of oxygen, neon, and iron are fixed to zero because we constrain their column densities through the {\sl slab} model. The abundances of the other elements are fixed to the solar values.

Dust is modeled with the {\sl amol} model in SPEX. This calculates the transmission of various molecules around the O and Si K-edge, and the Fe K / L-edges, using measured cross sections of various compounds taken from the literature (see the SPEX manual, Sect.~3.3). The most abundant interstellar compounds are included in this model. Currently, this model accounts for the XANES (X-ray Absorption Near Edge Structure) but not for the EXAFS (Extended X-ray Absorption Fine Structure).

Any evidence of an absorbing wind intrinsic to the binary system is modeled with additional {\sl slab} models in which we fit the velocity shift, but we fix the width to 50\,km\,s$^{-1}$. In summary, the model for the edge-by-edge analysis is defined as:
\begin{equation}
 F_{\rm empirical} = PL \cdot \prod_{i=1}^2S_{w,i} \cdot \prod_{i=1}^2S_{g,i} \cdot H \cdot \prod_{j=1}^4A_{j}.
 \label{eq:empirical_model}
\end{equation}

Here the powerlaw PL $(N,\Gamma)$ is absorbed by six components. Two partly-ionized {\sl slab} absorbers $S_{w,i\,}(N_{i},v_i,\sigma_i$) describe the wind. Two other slab absorbers $S_{g,i\,}(N_{i},v_i\equiv0,\sigma_i$) fit the low~/~intermediate-temperature (e.g. \ion{O}{i-v}) and high-temperature (e.g. \ion{O}{vi-viii}) ISM gas. 
The {\sl hot} model $H(N_{\rm H},T,A_{i},v_i\equiv0,\sigma_i)$ includes continuum neutral absorption, where $N_{\rm H}\equiv1.4\times10^{25}$m$^{-2}$, $T\equiv0.5$\,eV, and $A_{\rm O}\equiv\,A_{\rm Ne}\equiv\,A_{\rm Fe}\equiv~0$. Component~$A\,(N1,...,N4)$ includes up to four molecules and is characterized by the molecular indices and column densities.

\begin{figure}
\begin{center}
      \subfigure{ 
      \includegraphics[bb=65 115 510 755, width=6cm, angle=-90]{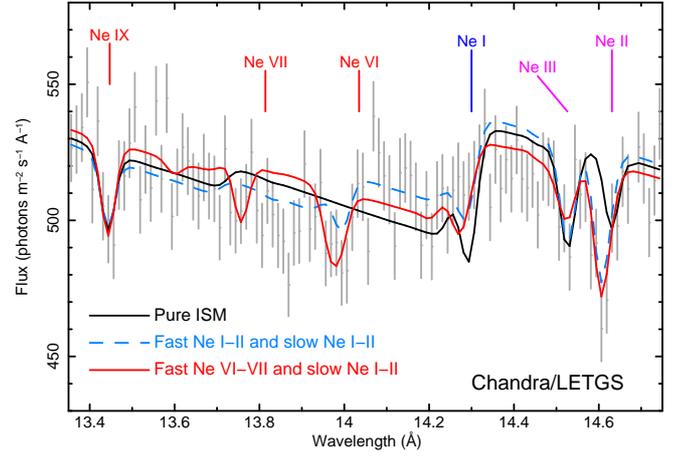}}
      \caption{\textit{Chandra}/LETGS data and alternative models near the Ne K edge. See also Table~\ref{table:neon}. Labels show the lines at their rest wavelengths. \ion{Ne}{iii} and \ion{Ne}{ix} lines are at rest and mostly produced by the ISM.}
          \label{fig:neon}
   \end{center}
\end{figure}

\subsection{The Ne K edge}
\label{sec:neon_edge}

In Fig.~\ref{fig:neon} we show the neutral absorption edges of neon as seen with LETGS. The \ion{Ne}{i} absorption is clearly located at 14.3\,{\AA}, but an equally-deep absorption feature is seen at 14.0\,{\AA}. Moreover, the \ion{Ne}{ii} absorption line at 14.63\,{\AA} appears to be blueshifted, while \ion{Ne}{iii} and \ion{Ne}{ix} are at rest. Note that in P13 the lower resolution of the RGS hampered a full description of the absorption in this spectral region (see the solid black line in Fig.~\ref{fig:neon} and the first model in Table~\ref{table:neon}).

We try three alternative models that imply absorption from outflowing gas (see Table~\ref{table:neon}). The features below 14.1\,{\AA} can be described as absorption by \ion{Ne}{i} or \ion{Ne}{vi-vii} in a fast outflow. The blueshift observed for the \ion{Ne}{ii} absorption line can be well modeled only with a slower ($\sim500$\,km\,s$^{-1}$) outflow of weakly ionized gas.

In principle, hot gas near the accretion disk recombines and produces emission lines that can be detected around the neon edge. 
Therefore, we have also tested an \ion{O}{viii} radiative recombination continuum ({\sl rrc} model in SPEX), which should be located at 14.23, but it did not improve the fit. 
This was expected since we do not detect any strong \ion{O}{viii} emission at 19.0\,{\AA}.

In Table~\ref{table:neon} we also give the column densities for the relevant ions fitted with these alternative models. We do not show the distribution of \ion{Ne}{iii} since it is mainly given by the gas at rest with a (log) column density of $20.0\pm0.1$ and is consistent within the different models.

\begin{table}
\caption{\textit{Chandra}/LETGS Ne\,K edge spectral fits (see Sect.~\ref{sec:neon_edge}).}  
\label{table:neon}    
\renewcommand{\arraystretch}{1.3}
\begin{center}
 \small\addtolength{\tabcolsep}{-4pt}
\scalebox{1}{%
\begin{tabular}{c c c c c c c}     
\hline                
\hline            
 Absorber                    & $\chi^2$/dof  & $v$ (km\,s$^{-1}$) & \ion{Ne}{i} &\ion{Ne}{ii} &\ion{Ne}{vi} &\ion{Ne}{vii}\,$^{(a)}$ \\  
\hline   
                  {\scriptsize ISM gas}        & $403/351$\,$^{(b)}$                     & $0$            & 21.4                 & 20.4  & $-$ & $-$  \\
\hline   
                  {\scriptsize ISM gas}    & \multirow{2}{*}{$378/348$}    & $0$            & 21.2                 & 20.2  & $-$ & $-$  \\
                  {\scriptsize Fast wind}    &                               & $-6450\pm250$  & 21.0                 & $<20$ & $-$ & $-$\\
\hline   
                  {\scriptsize ISM gas}      & \multirow{3}{*}{$353/347$\,$^{(b)}$}    & $0$            & 21.2                 & 19.2  & $-$ & $-$  \\
                  {\scriptsize Fast wind}      &                               & $-6350\pm250$  & 21.0                 & $<20$ & $-$ & $-$\\
                  {\scriptsize Slow wind}     &                               & $-520\pm150$   & 20.5                 & 20.5   & $-$ & $-$ \\
\hline   
 {\scriptsize ISM gas}      & \multirow{3}{*}{$345/347$\,$^{(b)}$}    & $0$            & 20.9                 & 19.8   & $-$ & $-$  \\
 {\scriptsize Fast wind}      &                               & $-1190\pm250$  & $-$                  & $-$  & 19.8 & 19.3\\
 {\scriptsize Slow wind}      &                               & $-540\pm150$   & 21.2                  & 20.6  & $-$ & $-$  \\
\hline                
                  {\scriptsize Total average}   &                   &           & {\scriptsize $21.3\pm0.1$}   & {\scriptsize $20.5\pm0.1$}  & $-$ & $-$ \\
\hline                
\end{tabular}}
\end{center}
$^{(a)}$ All column densities are in log (m$^{-2}$) with uncertainties of about 0.2. \\
$^{(b)}$ These are shown in Fig.~\ref{fig:neon}. Upper limits are given at the $2\,\sigma$ level.\end{table}

\subsection{The Fe L edge}
\label{sec:iron_edge}

\begin{figure}
  \begin{center}
      \subfigure{ 
      \includegraphics[bb=65 115 510 755, width=6cm, angle=-90]{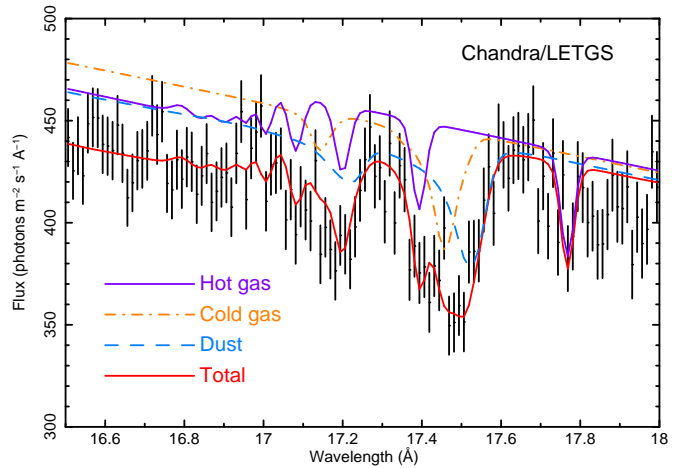}}
      \caption{\textit{Chandra}/LETGS Fe\,L edge (see also Table~\ref{table:iron}). The absorption of each component is also shown.}
          \label{fig:iron}
  \end{center}
\end{figure}

The iron L2 and L3 edges are located at 17.15 and 17.5\,{\AA} and show a rather complex structure (see Fig~\ref{fig:iron}). \ion{O}{vii} absorption lines are located at 17.4 and 17.77\,{\AA}. It is known that dust also affects the iron absorption edges (see e.g. \citealt{Lee2009}). As previously done in paper P13 for the RGS spectrum, we fit the LETGS iron edge by assuming the full ISM model consisting of dust, cold and hot gas. Neither velocity shift nor broadening is required, which means that most neutral Fe and \ion{O}{vii} originate in the ISM. The best fit of the 16--18.8\,{\AA} wavelength range is summarized in Table~\ref{table:iron}. The full model for the Fe\,L edge in the LETGS spectrum and each component are plotted in Fig.~\ref{fig:iron}. The best-fitting dust mixture consists of metallic iron (90\%) and hematite (Fe$_2$O$_3$, 10\%). The amount of \ion{O}{vii} was well constrained by including the \ion{O}{vii} 1s--3p (18.63\,{\AA}) absorption line. We did not jointly fit 
the \ion{O}{vii} 1s--2p here because near 21.6\,{\AA} there are additional lower ionization lines.

\begin{table}
\caption{\textit{Chandra}/LETGS Fe\,L edge spectral fits (see also Fig.~\ref{fig:iron}).}  
\label{table:iron}    
\renewcommand{\arraystretch}{1.3}
\begin{center}
 \small\addtolength{\tabcolsep}{-0pt}
\scalebox{1}{%
\begin{tabular}{c c c c}     
\hline                
\hline            
                  Absorber     & \ion{Fe}{i} $^{(a)}$ &\ion{O}{vii} $^{(a)}$ & $\chi^2$/dof               \\  
\hline   
                  Neutral gas  & 20.0                 & $-$                  & \multirow{3}{*}{$320/285$} \\
                  Ionized gas  & $-$                  & 21.2                 &                            \\
                  Dust         & 20.5                 & $-$                  &                            \\
\hline   
\end{tabular}}
\end{center}
$^{(a)}$ All column densities are in log (m$^{-2}$) with uncertainties of about 0.1.\end{table}

\subsection{The O K edge}
\label{sec:oxygen_edge}

\begin{figure}
  \begin{center}
      \subfigure{ 
      \includegraphics[bb=65 115 510 755, width=6cm, angle=-90]{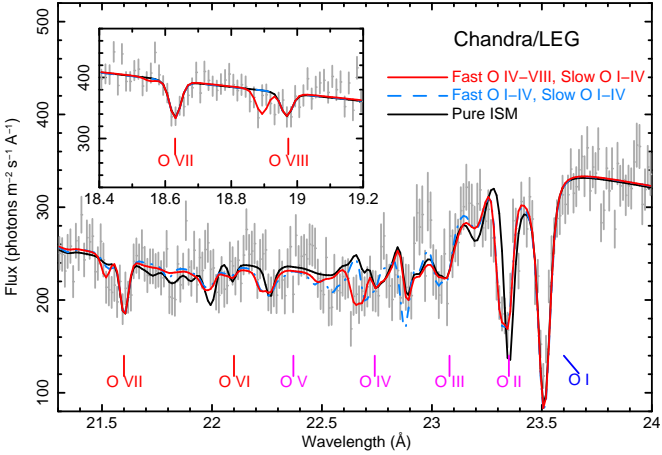}}
      \caption{\textit{Chandra}/LETGS data and alternative models near the O\,K edge (see Table~\ref{table:oxygen}). Labels show the lines at their rest wavelengths.}
          \label{fig:oxygen}
   \end{center}
\end{figure}

The spectral range between 21.5 and 23.5\,{\AA} is rich of absorption features produced by dust and all the atomic species from neutral \ion{O}{i} to \ion{O}{vii} (see Fig.~\ref{fig:oxygen}).
As previously done for the neon edge, we have tested different spectral models in order to constrain the significance of the wind. In Table~\ref{table:oxygen} we report three alternative models that have been applied to the $18-26$\,{\AA} spectral range.
The first model consists of pure ISM (gas + dust). The best-fit mixture of dust and molecules includes pyroxene MgSiO$_3$ (50\%), carbon monoxide CO (40\%), and hematite Fe$_2$O$_3$ (10\%). The column densities of hematite (on average $10^{20}$\,m$^{-2}$) as estimated with the oxygen and iron edges are consistent. Molecules like water ice or O$_2$ are not used since their abundances in the diffuse ISM are thought to be rather low (see e.g. \citealt{Jenkins2009}). Moreover, ice can be confused with pyroxene near the O\,K edge (see appendix in \citealt{Costantini2012}). 
The second model has two \ion{O}{i-iv} blueshifted {\sl slab} components with the velocity shifts fixed to those obtained for the neon edge with the two \ion{Ne}{i-ii} components (see also Table~\ref{table:neon}).
The third and best-fit model has two blueshifted components with different ionization states. 
In Fig.~\ref{fig:oxygen} we show the details of the \textit{Chandra}/LETGS O\,K edge with these three models and the positions of strongest lines at rest. The column densities of \ion{O}{v-vi} are not shown because we could only obtain upper limits.

\begin{table*}
\caption{Oxygen K edge spectral fits for the LETGS spectrum (see also Fig.~\ref{fig:oxygen} and Sect.~\ref{sec:oxygen_edge} for models detail).}  
\label{table:oxygen}    
\renewcommand{\arraystretch}{1.5}
\begin{center}
 \small\addtolength{\tabcolsep}{-2pt}
\scalebox{1}{%
\begin{tabular}{c c c c c c c c c}     
\hline                
\hline            
{Absorber}     &              $\chi^2$/dof           & $v$ (km\,s$^{-1}$) $^{(a)}$& {\ion{O}{i} $^{(b)}$} &{\ion{O}{ii}  $^{(b)}$} &{\ion{O}{iii} $^{(b)}$}   &{\ion{O}{iv} $^{(b)}$}  &{\ion{O}{vii} $^{(b)}$} &{\ion{O}{viii} $^{(b)}$}  \\  
\hline            
 ISM Gas       & \multirow{2}{*}{$861/626$}          & 0                 & $21.82\pm0.03$        & $21.3\pm0.1$           & $20.1\pm0.3$             & $20.1\pm0.2$           & $21.1\pm0.5$           & $19.7\pm0.2$        \\
 ISM Dust      &                                     & 0                 & $21.7\pm0.2$          & $-$                    & $-$                      & $-$                    & $-$                    & $-$    \\
\hline            
 ISM Gas       & \multirow{4}{*}{$744/616$}          & 0                 & $21.89\pm0.03$        & $20.7\pm0.1$           & $20.3\pm0.4$             & $19.8\pm0.5$           & $21.2\pm0.5$           & $19.7\pm0.2$   \\
 ISM Dust      &                                     & 0                 & $21.6\pm0.2$          & $-$                    & $-$                      & $-$                    & $-$                    & $-$       \\
 Slow wind     &                                     & $-520$            & $< 19.2$              & $20.6\pm0.1$           & $< 19.7$                 & $19.8\pm0.4$           & $-$                    & $-$       \\
 Fast wind     &                                     & $-6350$           & $< 20.3$              & $20.3\pm0.3$           & $19.9\pm0.5$             & $< 19.7$               & $-$                    & $-$       \\
\hline            
 ISM Gas       & \multirow{4}{*}{$727/616$}          & 0                 & $21.86\pm0.03$        & $20.8\pm0.1$           & $20.0\pm0.5$             & $20.0\pm0.2$           & $21.2\pm0.6$           & $19.7\pm0.2$  \\ 
 ISM Dust      &                                     & 0                 & $21.4\pm0.2$          & $-$                    & $-$                      & $-$                    & $-$                    & $-$      \\      
 Slow wind     &                                     & $-540$            & $< 19.8$              & $20.6\pm0.1$           & $< 19.8$                 & $< 19.8$               & $-$                    & $-$      \\      
 Fast wind     &                                     & $-1190$           & $-$                   & $-$                    & $< 19.8$                 & $20.0\pm0.3$           & $19.2\pm0.4$           & $19.5\pm0.2$  \\
\hline                
\end{tabular}}
\end{center}
$^{(a)}$ All velocity shifts are kept fixed. $^{(b)}$ All column densities are in log (m$^{-2}$). 
Upper limits are given at the $2\,\sigma$ level.\end{table*}

\section{Global spectral modeling}
\label{sec:spectral_models2}

The broadband spectral fits involve the simultaneous use of RGS and LETGS spectra. The MEG spectrum has been separately fitted for a comparison between different epochs. The spectral continuum between the RGS and LETGS epochs is different, while the interstellar absorption is thought not to change. Therefore, we have used the \textit{sectors} option in SPEX (for more detail see the SPEX manual and cookbook\footnote{http://www.sron.nl/spex/manual-hea-menu-1246.html}). This option allowed to keep the ISM parameters tied between all the observations while separately fitting the continuum of RGS and LETGS. At first, the parameters of the wind are tied between the two datasets in order to increase the statistics. Afterwards, we have unbound these parameters in order to search for long-term variability in the wind. The RGS\,1 and 2 data were treated as a single dataset.

\subsection{Model description}

For the simultaneous fit of the Mg, Ne, Fe, O, and N absorption edges we have edited the model in Eq.\,(\ref{eq:empirical_model}) by adopting a certain ionization balance for each gas component. The outflow is described with photoionized gas, which is provided by the {\sl xabs} model in SPEX. This model produces the transmission of a slab of photoionized material once given the hydrogen column density $N_{\rm H}$, the ionization parameter $\xi$, and the abundances $A_{i}$. 

The ISM modeling improves the multi-phase approach used in P12. The neutral gas is still described with the {\sl hot} model at lowest temperature, but we also fit the O, Ne, and Fe abundances that were fixed to zero during the edge-by-edge modeling. The warm phase is modeled with a combination of photoionized gas and collisionally-ionized gas. The former component accounts for the interstellar gas that is photoionized by hot stars radiation, while the latter includes intermediate ions such as \ion{O}{iv-v} that are thought to belong to a cooling phase lying between the warm and hot phases. The highly-ionized gas responsible for the \ion{O}{vi-viii} is described with another collisionally-ionized {\sl hot} component. Molecules and dust are still provided by the {\sl amol} model.

The continuum consists of a blackbody and a powerlaw component that reproduce the emission of the accretion disk and the surrounding corona. The \textit{physical} model is therefore defined as:
\begin{equation}
 F_{physical} = (PL + BB) \cdot \prod_{i=1}^2X_{w,i} \cdot \prod_{i=1}^3H_{i} \cdot X_{g} \cdot \prod_{j=1}^4A_{j},
  \label{eq:physical_model}
\end{equation}
where the powerlaw PL\,$(N,\Gamma)$ and the blackbody BB\,$(N,T)$ are absorbed by seven components.
Two photoionized {\sl xabs} absorbers $X_{w,i\,}(N_{\rm H,i},\xi_{i},A_{i},v_i,\sigma_i$) describe the wind. Another {\sl xabs} absorber $X_{g}$ provides the warm, photoionized interstellar gas (mostly \ion{O}{ii-iii} and \ion{Ne}{ii-iii}). Three {\sl hot} models $H_{i}$ mimic the cold (\ion{O}{i}, $T\equiv0.5$\,eV), warm-hot (\ion{O}{iv-v}), and hot (\ion{O}{vi-viii}, $T\sim0.1$\,keV) interstellar phases.

Abundances of the cold and hot gas are fitted since they may strongly deviate from the solar value due to dust depletion and recent stellar evolution, respectively, while those of the warm phases are fixed to solar. Component~$A$ (dust and molecules) is the same as the one used in the \textit{empirical} model.

The choice of the models as defined in Eqs.\,(\ref{eq:empirical_model}--\ref{eq:physical_model}) is dictated by the results obtained on the edge-by-edge modeling (see Sect.~\ref{sec:spectral_models1}), where we have shown alternative models that either provide worse fits or are difficult to justify.

\subsection{RGS--LETGS simultaneous fit}

We have jointly applied this global modeling to RGS\,1 and 2, and LETGS spectra in order to maximize the statistics. The LETGS spectrum dominates in the Mg, Ne, and Fe regions because of its higher resolution in this spectral range and the higher flux state of the source during the LETGS epoch. RGS leads the oxygen and nitrogen modeling due to its larger effective area. 

We notice that the physical models in SPEX already take into account thermal broadening and hence the fitted $\sigma_v$ is the turbulence component (square-summed to the thermal broadening). As previously done for the edge-by-edge modeling, we adopt $\sigma_v=10$\,km\,s$^{-1}$ for both the cold and warm ISM gas components, and $\sigma_v=50$\,km\,s$^{-1}$ for the winds (see Table~\ref{table:leg_rgs_fit}). The turbulence of the hot gas is left free, but we force it to range between 10 and $100$\,km\,s$^{-1}$. The parameters of the ISM absorbers are the same for the RGS and LETGS spectral models.

The full-band results obtained with the model in Eq.~(\ref{eq:physical_model}) are shown in Table~\ref{table:leg_rgs_fit}. The fully-joint fit is shown in the first column.
In this table we also show the significance expressed in terms of $\Delta\,\chi^{2}$ / degrees-of-freedom for each component in the joint model. The faster wind seems to have a higher ionization state.

Hydrogen only provides continuum absorption in the soft X-ray energy band. Therefore, absolute abundances in the wind are complicated to estimate. We set the value of the oxygen abundance equal to unity and used oxygen as reference atom. The hydrogen column density is the equivalent value obtained by the solar O/H value. All our fits suggested a high ($>5$) Ne/O abundance.

\begin{figure}
  \begin{center}
      \subfigure{ 
      \includegraphics[bb=21 73 535 477, width=0.64\textwidth, angle=-90]{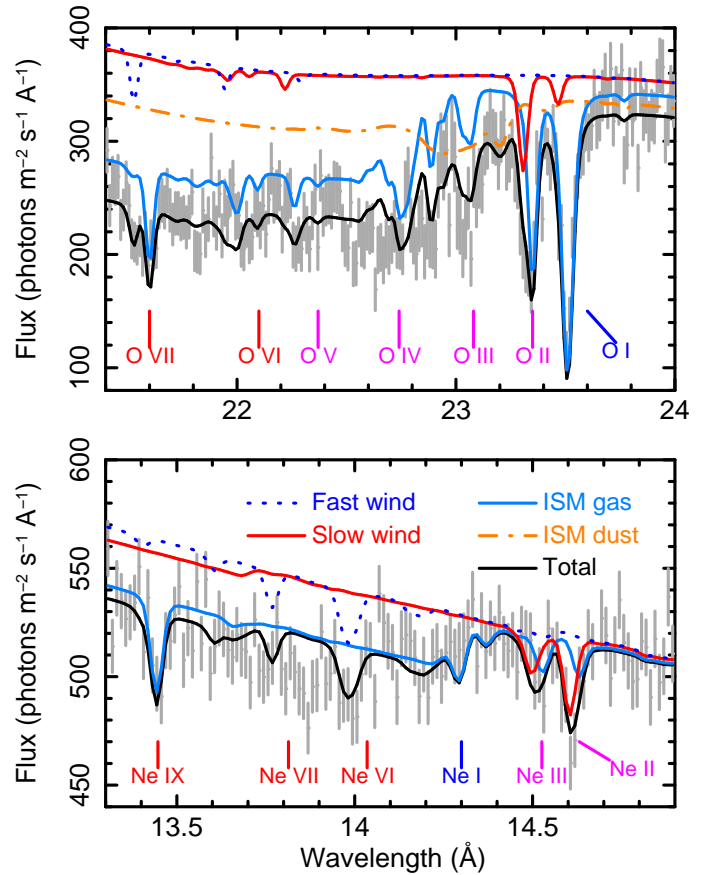}}
      \caption{\textit{Chandra}/LETGS data and joint RGS--LETGS model (see Table~\ref{table:leg_rgs_fit})
               near the oxygen (top panel) and neon (bottom panel) edges.
               The contributions of the main absorbers are also plotted.
               Dust is not shown near the Ne\,K edge because neon is not present in dust.
               Labels show the lines at their rest wavelengths.}
          \label{fig:LEG_semideco}
   \end{center}
\end{figure}

\begin{figure*}
  \begin{center}
      \subfigure{ 
      \includegraphics[bb=15 75 540 755, width=0.7\textwidth, angle=-90]{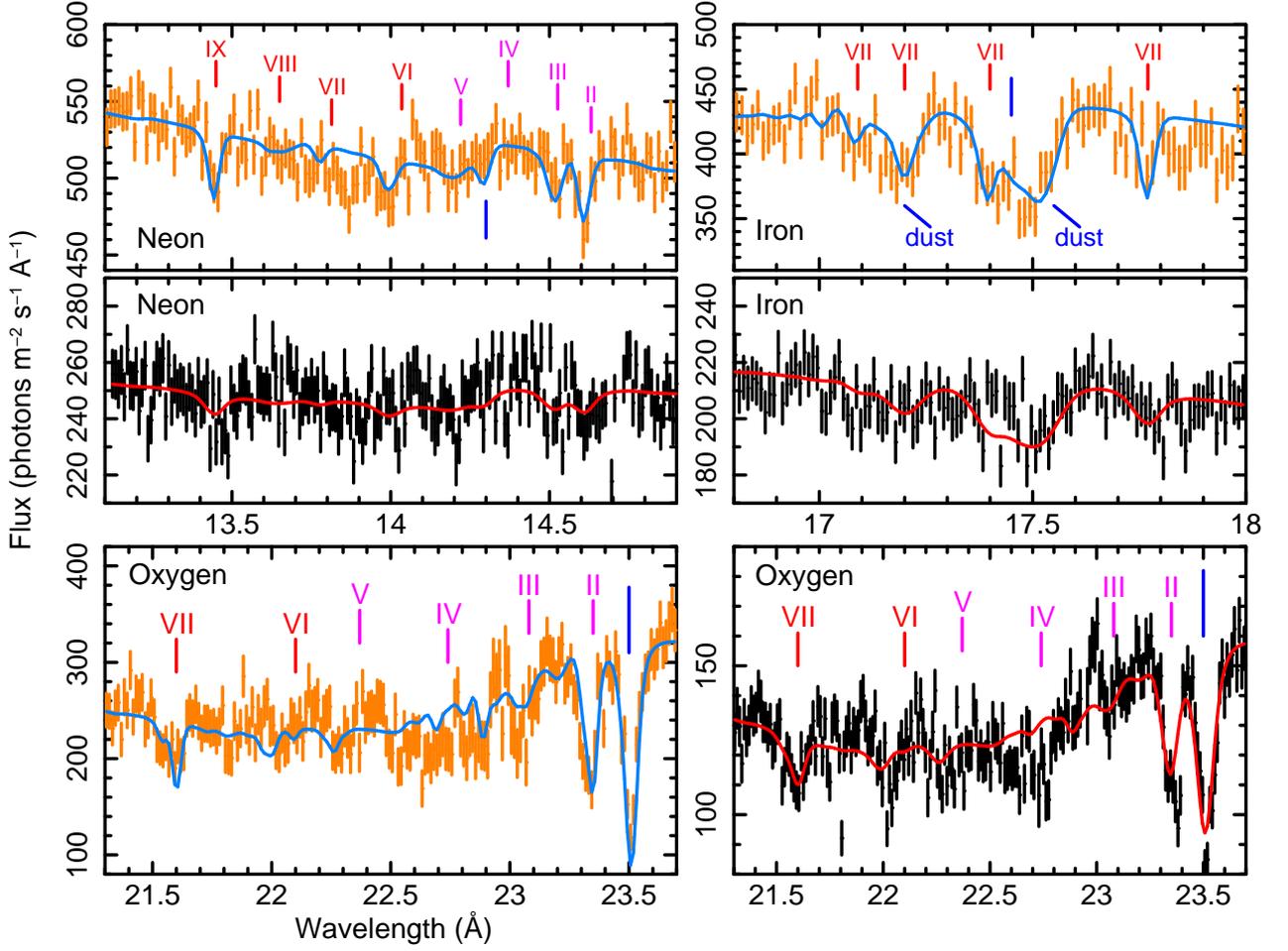}}
      \caption{Full-band joint RGS--LETGS fits (see also Table~\ref{table:leg_rgs_fit}, first column).
               Labels show the rest wavelengths and the ionization states of the lines. Blue marks provide the position of the neutral edges. Black points show the RGS spectrum, while the LETGS is in orange.}
          \label{fig:physical}
   \end{center}
\end{figure*}

\begin{table*}
\caption{Full-band spectral fits.}
\renewcommand{\arraystretch}{1.1}
\begin{center}
 \small\addtolength{\tabcolsep}{-3pt}
\scalebox{1}{%
\begin{tabular}{l|l|ll|ll|ll|l}
\hline                
\hline
Component             & Parameter                              & RGS  & LETGS     & RGS\,$^{(a)}$ & LETGS\,$^{(a)}$ & Pre-burst\,$^{(a,d)}$    & Post-burst\,$^{(a,d)}$ & MEG\,$^{(a)}$\\
\hline
\multirow{8}{*}{Cold gas} & $N_{\rm H}\,(10^{25} {\rm m}^{-2})$& \multicolumn{2}{c|}{$\equiv1.4$  }    & \multicolumn{5}{c}{$\equiv1.4$}  \\
                      & $kT$ (eV)                              & \multicolumn{2}{c|}{$\equiv0.5$  }    & \multicolumn{5}{c}{$\equiv0.5$}  \\
                      & $\sigma_V$ (km s$^{-1}$)               & \multicolumn{2}{c|}{$\equiv10$   }    & \multicolumn{5}{c}{$\equiv10$ }  \\
                      & N                                      & \multicolumn{2}{c|}{$1.0\pm0.2$}    & \multicolumn{5}{c}{$\equiv1$}    \\
                      & O                                      & \multicolumn{2}{c|}{$0.64\pm0.05$}    & \multicolumn{5}{c}{$\equiv0.64$} \\
                      & Ne                                     & \multicolumn{2}{c|}{$0.9\pm0.1$}    & \multicolumn{5}{c}{$\equiv0.9$} \\
                      & Mg                                     & \multicolumn{2}{c|}{$0.4\,(<1.3)$}    & \multicolumn{5}{c}{$\equiv0.4$}  \\
                      & Fe                                     & \multicolumn{2}{c|}{$0.1\,(<0.2)$}    & \multicolumn{5}{c}{$\equiv0.1$}  \\
                      & $\Delta\chi^2$/dof                     & \multicolumn{2}{c|}{24\,/\,5     }    & \multicolumn{5}{c}{ }            \\
\hline
\multirow{3}{*}{Warm gas 1} & $N_{\rm H}\,(10^{25} {\rm m}^{-2})$& \multicolumn{2}{c|}{$0.19\pm\,_{0.03}^{0.22}$}  & \multicolumn{5}{c}{$\equiv0.19$}  \\
                      & $\log 10^{-9} \xi$ (Wm)                  & \multicolumn{2}{c|}{$-2.94\pm0.24$} & \multicolumn{5}{c}{$\equiv-2.94$} \\
                      & $\sigma_V$ (km s$^{-1}$)                 & \multicolumn{2}{c|}{$\equiv10$  }   & \multicolumn{5}{c}{$\equiv10$}    \\
                      & $\Delta\chi^2$/dof                       & \multicolumn{2}{c|}{99\,/\,2   }    & \multicolumn{5}{c}{ }             \\
\hline
\multirow{3}{*}{Warm gas 2} & $N_{\rm H}\,(10^{25} {\rm m}^{-2})$& \multicolumn{2}{c|}{$0.05\pm0.01$}  & \multicolumn{5}{c}{$\equiv0.05$}  \\
                      & $kT$ (eV)                                & \multicolumn{2}{c|}{$12\pm2$   }    & \multicolumn{5}{c}{$\equiv12$}    \\
                      & $\sigma_V$ (km s$^{-1}$)                 & \multicolumn{2}{c|}{$\equiv10$}     & \multicolumn{5}{c}{$\equiv10$}    \\
                      & $\Delta\chi^2$/dof                       & \multicolumn{2}{c|}{76\,/\,2   }    & \multicolumn{5}{c}{ }             \\
\hline 
\multirow{9}{*}{Hot gas} & $N_{\rm H}\,(10^{25} {\rm m}^{-2})$ & \multicolumn{2}{c|}{$0.3\pm0.1$}    & \multicolumn{5}{c}{$\equiv0.3$}  \\
                      & $kT$ (eV)                              & \multicolumn{2}{c|}{$100\pm10$   }    & \multicolumn{5}{c}{$\equiv100$}   \\
                      & $\sigma_V$ (km s$^{-1}$)               & \multicolumn{2}{c|}{$10\,(<20)$  }    & \multicolumn{5}{c}{$\equiv10$}    \\
                      & C                                & \multicolumn{2}{c|}{$0.5\pm{0.3}$} & \multicolumn{5}{c}{$\equiv0.5$}   \\
                      & N                                      & \multicolumn{2}{c|}{$0.2\pm0.1$  }    & \multicolumn{5}{c}{$\equiv0.2$}   \\
                      & O                                      & \multicolumn{2}{c|}{$\equiv1$    }    & \multicolumn{5}{c}{$\equiv1$}     \\
                      & Ne                                     & \multicolumn{2}{c|}{$0.2\pm0.1$ }    & \multicolumn{5}{c}{$\equiv0.2$}   \\
                      & Mg                               & \multicolumn{2}{c|}{$0.5\pm{0.3}$} & \multicolumn{5}{c}{$\equiv0.5$}   \\
                      & Fe                               & \multicolumn{2}{c|}{$0.2\pm{0.1}$} & \multicolumn{5}{c}{$\equiv0.2$}   \\
                      & $\Delta\chi^2$/dof                     & \multicolumn{2}{c|}{306\,/\,8     }   & \multicolumn{5}{c}{ }             \\
\hline
\multirow{4}{*}{Dust $^{(b)}$}     & Pyroxene                  & \multicolumn{2}{c|}{$4\pm1$ }     & \multicolumn{5}{c}{$\equiv4$}     \\
                                   & Hematite                  & \multicolumn{2}{c|}{$1.0\pm0.5$}      & \multicolumn{5}{c}{$\equiv1$}     \\
                                   & Carbon monoxide           & \multicolumn{2}{c|}{$4\pm2$ }     & \multicolumn{5}{c}{$\equiv4$}   \\
                                   & Metallic iron             & \multicolumn{2}{c|}{$1.2\pm0.5$}      & \multicolumn{5}{c}{$\equiv1.2$}   \\
                      & $\Delta\chi^2$/dof                     & \multicolumn{2}{c|}{140\,/\,4 }       & \multicolumn{5}{c}{ }             \\
\hline
\multirow{5}{*}{Slow wind} &$N_{\rm H}\,(10^{25} {\rm m}^{-2})$& \multicolumn{2}{c|}{$0.04\pm0.01$ }  & $0.03\pm0.01$  & $0.05\pm0.01$      & $0.05\pm0.01$ & $0.04\pm0.01$         & $0.07\pm0.03$\\
                      & $\log 10^{-9} \xi$ (Wm)                & \multicolumn{2}{c|}{$-3.3\pm0.2$}    & $-3.0\pm0.5$   & $-3.3\pm0.2$       & $-3.3\pm0.2$  & $-3.6\pm0.3$          & $-3.0\pm0.4$ \\
                      & $\sigma_V$ (km s$^{-1}$)               & \multicolumn{2}{c|}{$\equiv50$}      & \multicolumn{2}{c|}{$\equiv50$}     & \multicolumn{2}{c|}{$\equiv50$} &        $\equiv50$  \\
                      & $V$ (km s$^{-1}$)                      & \multicolumn{2}{c|}{$-510\pm50$}     & \multicolumn{2}{c|}{$-570\pm40$}    & $-600\pm50$   & $-570\pm100$          & $-400\pm100$ \\
                      & Ne                                     & \multicolumn{2}{c|}{$13\pm6^{(c)}$}  & \multicolumn{2}{c|}{$10\pm4^{(c)}$} & \multicolumn{2}{c|}{$\equiv10$} &  $\equiv10$    \\
                      & $\Delta\chi^2$/dof                     & \multicolumn{2}{c|}{103\,/\,4}       & \multicolumn{2}{c|}{ }              &               &                       &              \\
\hline 
\multirow{5}{*}{Fast wind} &$N_{\rm H}\,(10^{25} {\rm m}^{-2})$&\multicolumn{2}{c|}{$0.005\pm0.002$}  & $0.007\pm0.003$   & $0.009\pm0.002$ & $0.006\pm0.002$ & $0.013\pm0.005$ & $0.018\pm0.008$\\
                      & $\log 10^{-9} \xi$ (Wm)                &\multicolumn{2}{c|}{ $+0.8\pm0.2$}    & $-0.2\pm0.3$      & $+0.9\pm0.1$    & $+0.9\pm0.3$ & $+0.5\pm0.5$       & $+1.2\pm{0.5}$\\
                      & $\sigma_V$ (km s$^{-1}$)               &\multicolumn{2}{c|}{ $\equiv50$}      & \multicolumn{2}{c|}{$\equiv50$}     & \multicolumn{2}{c|}{$\equiv50$}   & $\equiv50$     \\
                      & $V$ (km s$^{-1}$)                      &\multicolumn{2}{c|}{$-820\pm\,^{80}_{170}$}& \multicolumn{2}{c|}{$-1000\pm130$} & $-980\pm150$ & $-1100\pm150$         & $-1900\pm400$ \\
                      & Ne                                     &\multicolumn{2}{c|}{ $13\pm6^{(c)}$}  &   \multicolumn{2}{c|}{$10\pm4^{(c)}$} & \multicolumn{2}{c|}{$\equiv10$} &  $\equiv10$      \\
                      & $\Delta\chi^2$/dof                     &\multicolumn{2}{c|}{ 21\,/\,4}        & \multicolumn{2}{c|}{ }                &                &                               & \\
\hline 
\multirow{2}{*}{Blackbody}  & Flux (ph\,m$^{-2}$\,s$^{-1}$)\,$^{(e)}$ & $560\pm60$      &  $1330\pm70$      & $560\pm60$     & $1330\pm70$    & $1810\pm80$  & $1540\pm100$  & $1310\pm300$   \\
                            & $kT$ (eV)                               & $0.20\pm0.03$   &  $0.40\pm0.05$    & $0.20\pm0.03$  & $0.40\pm0.05$  & $0.40\pm0.05$ & $0.40\pm0.05$ & $0.09\pm0.02$\\
\multirow{2}{*}{Powerlaw}   & Flux (ph\,m$^{-2}$\,s$^{-1}$)\,$^{(e)}$ & $3850\pm40$     &  $7490\pm150$     & $3850\pm40$    & $7490\pm150$   & $7065\pm150$  & $7080\pm220$  & $3620\pm60$   \\
                            & $\Gamma$                                & $2.1\pm0.1$     &  $2.3\pm0.1$      & $2.1\pm0.1$    & $2.3\pm0.1$    & $2.3\pm0.1$   & $2.3\pm0.1$   & $2.1\pm0.1  $\\
\hline 
\multirow{1}{*}{Statistics} & $\chi^2$ / dof                   & \multicolumn{2}{c|}{8841/5838}       & \multicolumn{2}{c|}{8828/5834}  & 2010/1574 & 1864/1574 & 815/1242\\
\hline
\end{tabular}}
\label{table:leg_rgs_fit}
\end{center}
$^{(a)}$ Semi-joint fits with wind parameters decoupled. $^{(b)}$ Dust column densities are in units of $10^{20}\,{\rm m}^{-2}$. $^{(c)}$ Ne abundances are coupled. \\
$^{(d)}$ Pre- and post-burst LETGS spectra. 
$^{(e)}$ Fluxes are calculated in the 7--27\,{\AA} spectral range.
All upper limits are given at the $2\,\sigma$ level.
\end{table*}

\subsection{Limitations to the models}
\label{sec:limitations}

There are some factors that may introduce systematic errors in our best-fitting models. First of all, the features produced by molecules near the iron and oxygen edges are similar and degenerate. These were already discussed in \citealt{Costantini2012} and in P13. We refer to their papers for a thorough discussion of the problems concerning the molecular models. The full-band fits break as much as possible the degeneracy.

Another problem is the presence of several bad pixels or dead columns in the RGS spectra that may affect the estimates of some parameters. This is solved by simultaneously fitting RGS and LETGS spectra, but the correction is only partial if spectral line variability occurs.

Finally, in the modeling of the photoionized absorbers we have used the default ionization balance and spectral energy distribution (SED) in SPEX because there is no simultaneous IR-to-UV coverage of the source. The default SED in SPEX refers to AGN NGC\,5548, which certainly differs from that of an X-ray binary like SAX\,J1808 or from the ISM SEDs. We have tested the SEDs given in P12 that are produced by starlight, local galaxies and QSOs, only local QSO, and cosmic background radiation plus X-ray background. The fits were comparable and only the ionization parameter significantly scatters, {which is expected since the ionization balance differs between the SEDs (see Table~\ref{table:sed_tests}). A better modeling of the SED may slightly improve the fit, but this is beyond the aim of this paper.} We do not show the other model parameters because they were consistent within the errors.

\begin{table}
\caption{Full-band (RGS$\,+\,$LETGS) spectral fits with different SEDs.}  
\label{table:sed_tests}    
\renewcommand{\arraystretch}{1.5}
\begin{center}
 \small\addtolength{\tabcolsep}{-2pt}
\scalebox{1}{%
\begin{tabular}{c| c c c c c c}     
\hline                
\hline            
 Par  $^{(a)}$                & Default      & Starlight    & LE           & QSO          & BKG          \\  
\hline   
       $\xi_{\rm ISM}$        & $-2.9\pm0.2$ & $-1.5\pm0.1$ & $-2.8\pm0.5$ & $-2.2\pm0.4$ & $-2.8\pm0.2$ \\  
       $\xi_{\rm wind1}$      & $-3.3\pm0.2$ & $-1.3\pm0.2$ & $-2.1\pm0.2$ & $-1.8\pm0.1$ & $-1.7\pm0.1$ \\  
       $\xi_{\rm wind2}$      & $+0.8\pm0.2$ & $>0.7$       & $+1.5\pm0.1$ & $+1.4\pm0.1$ & $+1.3\pm0.3$ \\  
\hline   
       $\Delta\chi^2$         & $-$          &  $+24$       & $-9$         & $-7$         & $-11$        \\  
\hline   
\end{tabular}}
\end{center}
$^{(a)}$ All ionization parameters are in units of $\log 10^{-9} \xi$ (Wm). \end{table}

\section{Variability}
\label{sec:variability}

\subsection{Long--term variability}

The powerlaw and blackbody fluxes doubled during the LETGS epoch, which may potentially lead to some variability in the winds if they originate near the accretion disk. Therefore, we have performed additional fits by decoupling the column densities and ionization states of the photoionized outflowing absorbers between RGS and LETGS spectra. The kinematics are still coupled because RGS has a lower spectral resolution. We fixed all the ISM parameters to those obtained with the fully-joint fit and show the results in Table~\ref{table:leg_rgs_fit}, second column. The slow wind component does not change significantly.

Finally, we separately fitted the archival \textit{Chandra}/MEG spectrum within $6-25$\,{\AA}. The two wind components are still detected at a level higher than $2\sigma$ and are consistent with the LETGS epoch, but the uncertainties are larger due to the short exposure.

\subsection{Burst-induced variability}

ZA13 reported that the SAX\,J1808 thermonuclear burst occurred during the LETGS epoch was the brightest ever observed with \textit{Chandra}. They looked for burst-related spectral features by extracting the spectrum during the burst event. Here, we search for any burst-related signature by comparing the integrated spectra extracted by cutting the event file in two (pre- and post-burst) event files in order to maximize the statistics (see Sect.~\ref{sec:data}).

We have fitted the pre- and post-burst spectra and reported the results in Table~\ref{table:leg_rgs_fit}, third column. There is no significant variability in the wind during the LETGS epoch. The blackbody flux instead is slightly decreased after the burst.

\subsubsection{A narrow emission-line component?}
\label{sec:emission_lines}

\begin{figure}
  \begin{center}
      \subfigure{ 
      \includegraphics[bb=65 115 510 755, width=6cm, angle=-90]{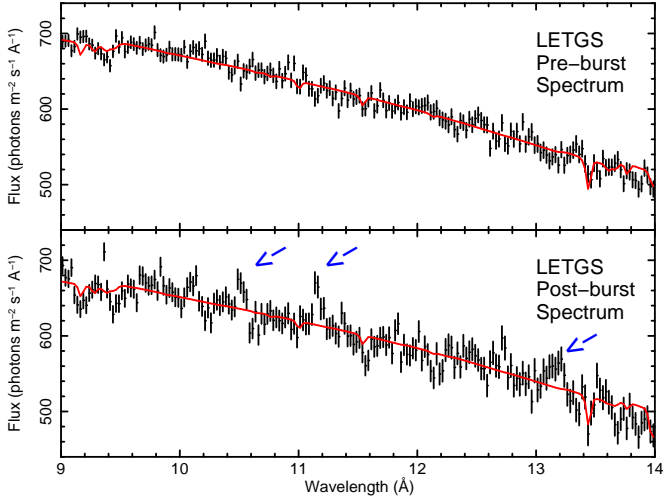}}
      \caption{{LETGS spectra before and after/including the burst events together with their models as given in Table~\ref{table:leg_rgs_fit}.
               Note the emission-like features indicated by the blue arrows in the post-burst spectrum. 
               Data have been binned by a factor of 2 for displaying purposes.}}
          \label{fig:LETGS_pre_and_post_burst}
   \end{center}
\end{figure}

\begin{figure}
  \begin{center}
      \subfigure{ 
      \includegraphics[bb=65 115 510 755, width=6cm, angle=-90]{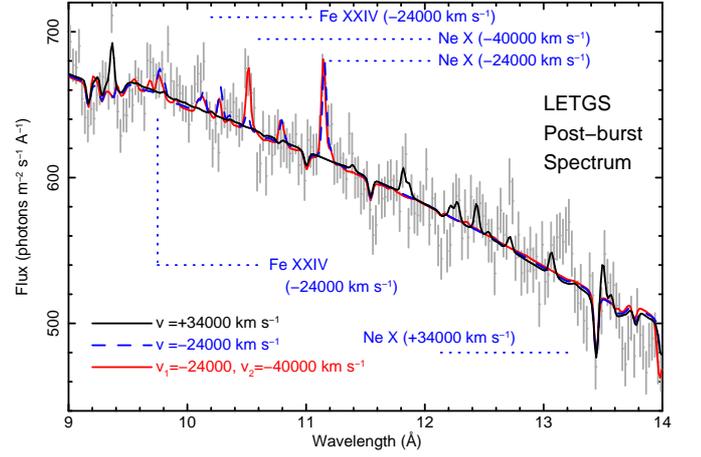}}
      \caption{LETGS spectrum after (and including) the burst events together with three alternative CIE models for the emission lines.
               The dashed blue lines show the shifts from the rest-frame wavelengths.
               Data have been binned by a factor of 2 for displaying purposes.}
          \label{fig:LETGS_postburst}
   \end{center}
\end{figure}

On average, SAX\,J1808 was slightly brighter before the burst. The ratio between the post-burst and pre-burst spectra is generally within or below unity throughout the $7-27$\,{\AA} band. However, there are a few exceptions that are represented by some emission-like features in the post-burst spectrum between 9 and 14\,{\AA} (see Fig.~\ref{fig:LETGS_pre_and_post_burst}--\ref{fig:LETGS_postburst}). The strongest peaks are at 10.5 and 11.1\,{\AA}.

We tentatively fitted these emission-like features with three alternative models involving collisionally-ionized emission (CIE) components (\textit{cie} model in SPEX). We show the free parameters as well as the changes in the $\chi^2$ and the additional degrees of freedom in Table~\ref{table:emission_lines}. These components reproduce well the narrow emission lines, see Fig.~\ref{fig:LETGS_postburst}, but they require high velocity shifts and high neon abundances.

\begin{figure}
  \begin{center}
      \subfigure{ 
      \includegraphics[bb=65 110 510 755, width=6cm, angle=-90]{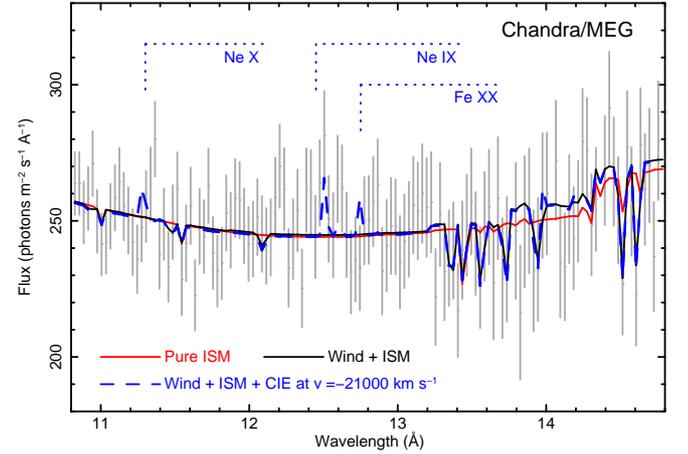}}
      \caption{Detail of the MEG spectrum near the Ne\,K edge with three alternative models. 
               The solid black line shows our standard model (see Table~\ref{table:leg_rgs_fit}).
               The blue dashed line model includes a blueshifted CIE emission.
               The shifts from the rest-frame wavelengths are shown like in Fig.\ref{fig:LETGS_postburst}.}
          \label{fig:MEG_models}
   \end{center}
\end{figure}

Similar features are seen in the MEG spectrum, but the low S/N ratio due to the short exposure forbids to put relevant constraints and only one, colder \textit{cie} component ($T<0.6$\,keV and $v=-(21\pm3)\times10^3$km\,s$^{-1}$) is marginally detected at a 1.5$\sigma$ level (see Fig.~\ref{fig:MEG_models}).

\begin{table}
\caption{Emission-line spectral fits of the LETGS spectrum after (and including) the burst events (see also Fig.~\ref{fig:LETGS_postburst}).}  
\label{table:emission_lines}    
\renewcommand{\arraystretch}{1.3}
\begin{center}
 \small\addtolength{\tabcolsep}{-1pt}
\scalebox{1}{%
\begin{tabular}{c| c c c c c}     
\hline                
\hline            
 Model                        & $\Delta\chi^2$/dof      & $v$ ($10^3$km\,s$^{-1}$) & $n_{\rm e} n_{\rm X} V$\,$^{(a)}$  & T (keV) & Ne/O\,$^{(b)}$ \\  
\hline   
       { 1 CIE}               & $24/5$                  & $+34\pm1$                & $3\pm1$                  & $2.3\pm\,_{0.7}^{1.3}$ & $>2$    \\
\hline 
       { 1 CIE}               & $39/5$                  & $-24\pm1$                & $4\pm2$                  & $2.5\pm\,1.0$          & $>2.5$  \\
\hline 
       \multirow{2}{*}{2 CIE} & \multirow{2}{*}{$60/5$} & $-24\pm1$                & $3\pm\,^{3}_{1}$         & $1.9\pm\,_{0.5}^{0.9}$ & \multirow{2}{*}{$>3$} \\
                              &                         & $-40\pm1$                & $2.4\pm\,^{1.6}_{0.9}$   & $2.8\pm\,_{0.6}^{1.7}$ &              \\
\hline   
\end{tabular}}
\end{center}
$^{(a)}$ Emission measures $n_{\rm e} n_{\rm X} V$ are in units of ($10^{62}$m$^{-3}$). \\
$^{(b)}$ Ne/O abundance ratios are in solar units. \end{table}

\section{Discussion}
\label{sec:discussion}

We have analyzed the high-resolution X-ray spectra of the LMXB SAX~J1808 taken with the gratings on board of XMM-\textit{Newton} (RGS) and \textit{Chandra} (LETGS and MEG), which span a period of about six years (see Table~\ref{table:1}). We try to understand whether the high neon abundance in the ISM as estimated in P13 (and previously by \citealt{Patruno2009}) was due to interstellar neon enrichment or to some metal-rich absorbing media located near the source.

\subsection{The source continuum}
\label{sec:discussion_continuum}

The spectral continuum was described with a combination of blackbody and powerlaw emissions which reproduce the flux originating from the accretion disk and its corona. 

The continuum shape has radically changed between the three epochs, which was expected given the transient nature of the source. The disk (blackbody) emission decreased from 2005 (MEG) to 2008 (RGS), while during 2011 (LETGS) it came back to the early level (see Table~\ref{table:leg_rgs_fit}). 

The corona (powerlaw) has brightened from 2005 to 2011 and the spectrum has generally hardened. The gas surrounding the binary system may have been dynamically affected by the variability of this strong radiation field (see also Sect.~\ref{sec:discussion_wind_variability}).

\subsection{The wind}

We have accurately analyzed the neon absorption edge in Sect.~\ref{sec:neon_edge} and detected several absorption features that were described with alternative absorption models implying blueshifted plasmas at different ionization states (see also Fig.~\ref{fig:neon}). All models provide acceptable fits, but some of them are less feasible. Absorption by fast neutral neon was also used by \citet{Schulz2010} to describe the dynamical neon edge of the UCB 4U\,0614+091, but they noticed that it is difficult to produce a large amount of neutral neon in a hot accretion disk. Higher ions should be then detected at similar velocities.

The most reasonable description of the neon edge implies absorption by outflowing gas with velocities of $500-1000$\,km\,s$^{-1}$ and a large range of ionization states like \ion{Ne}{ii-vii}, which is confirmed by several features produced by \ion{O}{ii-vii} outflowing at similar velocities (see Figs.~\ref{fig:oxygen}$-$\ref{fig:LEG_semideco}).

According to our modeling, this outflowing gas should be in photoionization equilibrium, but there seem to be (at least) two different dynamical states: a faster plasma ($1000$\,km\,s$^{-1}$) with higher ionization (\ion{Ne}{vi-vii}) and a slower one ($500$\,km\,s$^{-1}$) with lower ionization (\ion{Ne}{ii-iii}). If the wind originates from the accretion disk atmosphere, hence the two states may have risen from different regions / altitudes therein.

Outflows with intermediate ionization states {(e.g. \ion{O}{ii-iv})} and velocities similar to ours have already been detected in other LMXBs (see e.g. \citealt{Ioannou2003} and \citealt{Costantini2012}). {However, outflowing gas is more often observed at a higher ionization state} (see e.g. \citealt{Miller2006}). These winds are generally thought to be launched by thermal and/or radiation pressure near the accretion disk atmosphere (for a review, see \citealt{DiazTrigo2012} and references therein).

\subsubsection{A continuum wind atmosphere ?}
\label{sec:discussion_warm_AMD}

It is possible to check if the wind is discrete or continuous in the $\xi$ parameter space through the \textit{warm} model in SPEX, which is a multi-component version of the \textit{xabs} model. This model consists of a continuous distribution of column density $N_{\rm H}$ as a function of the ionization parameter $\xi$ with the same outflow velocity and turbulent broadening. The model assumes a logarithmic grid of equidistant values of log\,$\xi$ and determines the value of $f_i=d\,N_{\rm H}\,/\,d\,\log\,\xi_i$ by making a cubic-spline interpolation in the $f_i-\log\xi$ space (consult the SPEX manual for more detail).

\begin{figure}
  \begin{center}
      \subfigure{ 
      \includegraphics[bb=68 38 522 724, width=6cm, angle=+90]{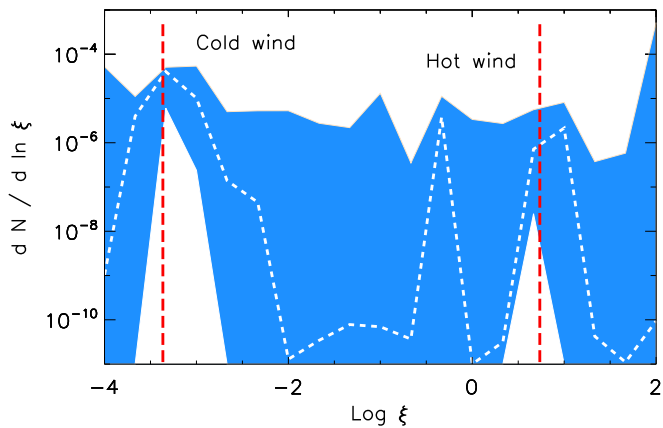}}
      \caption{Absorption measure distribution of the photoionized wind (joint RGS--LETGS \textit{warm} model).
      The red lines show the two $\xi$ values obtained with our standard model (see Table~\ref{table:leg_rgs_fit}, first column).}
          \label{fig:SPEX_warm_AMD}
   \end{center}
\end{figure}

We have thus substituted the two (wind) \textit{xabs} with one \textit{warm} model and simultaneously fitted the RGS--LETGS spectra again as previously shown in the first column of Table~\ref{table:leg_rgs_fit}. The goodness of the fit is comparable and the new ISM parameters are fully consistent with those obtained with the standard model. The absorption measure distribution (AMD) of the \textit{warm} model is shown in Fig.~\ref{fig:SPEX_warm_AMD}, where the white dashed line gives the best fit, the blue area provides the $1\sigma$ contours, and the vertical red dashed lines refer to the two best-fitting $\xi$ values from our standard \textit{xabs} model. On average, the multi-ionization gas has v\,$=-540\pm30$\,km\,s$^{-1}$ and Ne/O\,$=7\pm4$ abundance ratio, which are strongly driven by the slower wind due to its higher column density (see Table~\ref{table:leg_rgs_fit}).

The AMD modeling confirms the presence of at least two ionization ranges $(-3.5,-3.0)$ and $(+0.5,+1.0)$, which are fully consistent with the two best-fit $\xi$ values ($-3.3$ and $+0.8$) from our standard model, but we cannot entirely exclude a continuous, non-monotonic $\xi$ distribution. Additional FUV spectra (unfortunately unavailable for this source) may be necessary to obtain more constraints on the wind $\xi$ distribution.

\subsubsection{Wind variability}
\label{sec:discussion_wind_variability}

The spectral continuum of SAX\,J1808 radically changed in the six years between the three observations (see Sect.~\ref{sec:discussion_continuum}). We have searched for variability in the wind absorbers by decoupling their parameters in the simultaneous spectral fits and comparing them to the continuum changes (see Table~\ref{table:leg_rgs_fit}).

The slow component is generally stable within the three epochs. The fast component instead has a lower ionization in the XMM-\textit{Newton} observation, during which the fluxes of both the accretion disk (blackbody) and the corona (powerlaw) were lower. The \textit{Chandra}/MEG epoch is characterized by a highly ionized fast wind, a weaker corona and a brighter accretion disk. The wind parameters do not change between the pre- and post-burst LETGS epochs through which the continuum was mostly stable (see the blackbody and the powerlaw in Table~\ref{table:leg_rgs_fit}).

It is difficult to understand whether and how the ionization state of the fast wind is affected by the continuum, but it seems to depend on the flux of the accretion disk (blackbody). This would suggest an accretion disk origin for the fast wind, at least.

\subsection{Further insights on the binary system}

We obtained velocities within 500$-$1000\,km\,s$^{-1}$ for the wind. These values are just above the 300$-$400\,km\,s$^{-1}$ value estimated for the companion by \citet{Elebert2009} through observations of the \ion{N}{iii} $\lambda4640.64$ Bowen blend. Moreover, they also reported that the H\,$\beta-\gamma$ Balmer absorption lines appear to be significantly blueshifted. This also suggests a disk origin for the wind.

Winds are normally seen at high $(\gtrsim45^{\circ})$ inclinations where the gas density may be high and the ionization not too high (see e.g. \citealt{DiazTrigo2012} and references therein). This agrees with the values of $50^{\circ}-60^{\circ}$ found for SAX\,J1808 in the literature (see e.g. \citealt{Deloye2008}) and further supports the accretion disk wind. 

All our wind models show a high Ne/O abundance ratio. It is difficult to determine whether this is due to a neon enrichment or to an oxygen depletion in the accretion disk. UV spectra may provide the amount of hydrogen in the outflow.

Recent work dedicated to the measure of the binary system masses came out with a broad range of values: $M_{\rm NS} = 0.4-2.2\,M_{\odot}$ and $M_{\rm donor} = 0.04-0.11\,M_{\odot}$ (see e.g. \citealt{Bildsten2001}, \citealt{Deloye2008D}, and \citealt{Wang2013}). The exact values strongly depend on the inclination and distance adopted for the system. However, most authors suggest a brown dwarf companion for SAX\,J1808.4.

Brown dwarf atmospheres are thought to have subsolar abundances due to the low stellar mass (see e.g. \citealt{Baraffe1998} and references therein), but since the binary system is at a small distance, we may not expect large deviation from Solar abundances. This anyway disagrees with our high Ne/O estimates. One reason may be that the companion is a white (rather than a brown) dwarf and has indeed a high neon abundance. Alternatively, the companion may be a brown dwarf and its envelope is significantly exhausted by the accretion and what we currently see is gas extracted from the stellar interior. Finally, if the NS mass is not too high, bursts may have expelled some metal-rich material from its envelope and spread it in the surrounding environment.

It is complicated to obtain more accurate results especially due to the faintness of the companion even in the optical band. A combined FUV~/~X-ray observation with the high-resolution spectrometers on board of HST and \textit{Chandra} may constrain the chemistry of the accretion disk and provide more insights on the nature of the binary system.

\subsection{Alternatives to the wind ?}

As previously mentioned, \citet{Schulz2010} found a similar modulation of the neon K\,edge in the \textit{Chandra} spectrum of the LMXB 4U\,0614$+$091. They also warned about the feasibility of the very fast wind of neutral neon and proposed a further description involving Doppler-shifted \ion{O}{viii} emission. This was supported by the detection of a broad emission-like feature between 17 and 19\,{\AA}, which can be identified as a relativistically broadened \ion{O}{viii} Ly{$\alpha$} line (see also \citealt{Madej2010}). However, the interpretation of the neon edge in terms of \ion{O}{viii} radiative recombination continuum (RRC) required very high velocities and an additional (but not detected) emission line at lower velocities.

We have tested an RRC emission model for the neon edge in the SAX\,J1808 LETGS spectrum without obtaining interesting results (see Sect.~\ref{sec:neon_edge}) and we also noticed that no strong \ion{O}{viii} emission is detected around 19\,{\AA} (see e.g. Fig.~\ref{fig:oxygen}). Another description of the neon edge may be yet possible, but it is very complicated to find an alternative explanation for the blueshifted (and somewhat asymmetric) \ion{O}{ii-vii} absorption lines (see Fig.~\ref{fig:oxygen}).

\subsection{Line--emitting plasma}

We have detected a set of narrow emission lines between 10--14\,{\AA} in the LETGS spectrum which includes the event during and after the bright burst (see Sect.~\ref{sec:emission_lines}). We tentatively modeled them with a single or double CIE component. These require a Ne/O abundance as high as in the wind, {but velocities and ionization states higher than the wind by more than an order of magnitude (see Table~\ref{table:emission_lines}).} This suggests that the emitting outflow should originate presumably closer to the NS. {The Doppler shifts are within $(20-40)\times10^3$\,km\,s$^{-1}$, or $\sim0.1c$, and are larger than what has been observed in other systems.} \citealt{Iaria2013} also analyzed similar high-ionization emission lines in the XMM-\textit{Newton} and \textit{Chandra} grating spectra of the LMXB X1822--371, but their Doppler shifts were below 1000\,km\,s$^{-1}$, and proposed an origin within the accretion disk.

ZA13 noticed that the SAX\,J1808 burst is Eddington-limited and therefore could have a wind expelling heavy elements from the NS atmosphere. Expulsion velocities may be up to 0.01$c$ \citep{Paczynski1986}. Very extreme cases with velocities of about 0.1$c$ have been observed but in more energetic bursts \citep{zand2010}.

{If the detection of both redshifted and blueshifted emission is real then it could imply a jet originating near the neutron star and possibly related to its magnetic field. However, the kinematics of the redshifted and blueshifted emitters are different and there is no evidence for relativistical-broadening, which makes this interpretation difficult. The narrowness of the lines rules out possible effects due to the rotation of the system, which would suggest a limited time for the emission.}

We could not significantly detect the emission lines in the pre-burst LETGS spectrum {(see Fig.\ref{fig:LETGS_pre_and_post_burst}, top panel)}. These cannot be detected in the RGS, but we notice that the RGS spectral resolution is lower than that of LETGS below 14\,{\AA}. The MEG spectrum has a much lower S/N compared with the LETGS spectrum and we only found a 1.5$\sigma$ detection, which can also be attributed to spectral noise. However, we noticed that this observation was taken one year after an outburst and it was proposed in order to search for intrinsic spectral features \citep{Galloway2005A}. A deep observation taken with the \textit{Chandra} High-Energy Transmission Grating Spectrometer (HETGS) could determine the nature of the emission line and indicate whether they are burst-related.

\subsection{The diffuse interstellar medium}

The ISM as seen in the RGS spectra of SAX\,J1808 (and other eight LMXBs) was extensively studied and described in P13. {Therefore, we will briefly discuss our new results on the ISM.}

\subsubsection{Interstellar phases}

In P13 the interstellar medium was decomposed into multi-temperature gas, dust, and molecules. Here we confirm their model. The warm intermediate-ionization gas (\ion{O}{ii-v}) cannot be described by a single photo- or collisionally-ionized gas component (see Table~\ref{table:leg_rgs_fit}). The cold (\ion{O}{i}) and hot (\ion{O}{vi-viii}) phases can instead be well described by one isothermal component each.

\subsubsection{Interstellar chemistry}

The total abundances of the cold phase are estimated by summing the contributions to the column densities from the neutral gas and dust, and calculating their ratios with the column densities expected from the solar values (once adopted a hydrogen column density of $1.4\times 10^{25}$\,m$^{-2}$, see also Sect.~\ref{sec:data}).

The total abundances are shown in Table~\ref{table:total_abundances} together with the percentage contributions of gas and dust. The dust depletion factors for O, Mg, and Fe are consistent with those found in the literature (see e.g. \citealt{Jenkins2009} and references therein). Neon is a noble gas and has no depletion. In our standard model we did not include nitrogen molecules because we could not detect any significant feature from molecules like N$_2$O and because nitrogen depletion is also thought to be low.

The ISM total abundances are fully consistent with the solar values of \citet{Lodders09}, which is expected because SAX\,J1808 ($l=25^{\circ}$, $d\sim3$\,kpc) should be at about 7\,kpc from the Galactic center. The LOS towards the source does not cross the inner regions of the Galaxy and therefore the ISM abundances are not strongly affected by the metallicity gradient.

The reason why our updated value for the neon abundance agrees with that of the Solar System is due to the subtraction of the absorption contribution due to material intrinsic to the binary system. In principle, the search for CSM absorption could have been done in our previous work, but the lower resolution and the shorter exposure of the available spectra made it difficult.

The (new) ISM abundances also support the wind model. An alternative approach in terms of emission lines for the neon and oxygen edges would indeed require additional absorption at rest and unexpectedly high neon (1.5) and oxygen (1.3) abundances.

The Mg abundance for SAX\,J1808.4 as estimated in P13 was subsolar. This was probably due to the degeneracy introduced by using molecules that have similar absorption edges such as pyroxene silicates and water ice. Here we prefer not to use water ice because it is thought to be rare in the diffuse ISM and in order to avoid the degeneracy. Moreover, the new deep \textit{Chandra}/LETGS spectrum provided us with higher spectral resolution and S/N ratio necessary for the modeling of the Mg K\,edge at 9.5\,{\AA}.

\begin{table}
\caption{Total abundances and fractionation of the cold neutral phase.}  
\label{table:total_abundances}    
\renewcommand{\arraystretch}{1.3}
\begin{center}
 \small\addtolength{\tabcolsep}{-0pt}
\scalebox{1}{%
\begin{tabular}{c c c c}     
\hline            
         Element & Abundance $^{(a)}$  & \% in gas & \% in dust \\  
\hline   
         N       & $1.0\pm0.2$         &   100     &     0      \\
         O       & $0.9\pm0.1$         &  70--85   &    15--30  \\
         Ne      & $0.9\pm0.1$         &   100     &     0      \\
         Mg      & $1.0\pm0.3$         &  0--50    &    50--100 \\
         Fe      & $0.8\pm0.2$         &  0--25    &    75--100 \\
\hline   
\end{tabular}}
\end{center}
$^{(a)}$ Abundances are units of \citet{Lodders09}.\end{table}

\section{Conclusion}
\label{sec:conclusion}

In this paper we have presented a detailed analysis of the soft X-ray spectra at different epochs of the LMXB SAX\,J1808.4-3658. We have carefully measured the abundances of the ISM which lies in the LOS towards this source. They all agree with the abundances of the Solar System as expected due to the nearness of the source.

In this paper we took into account not only the cold ISM phase, but we
also looked for absorption by gas in the source system. A better
modeling of the absorption edges region allowed us a precise estimates of
elemental abundances. We found an ISM Ne abundance consistent with the Solar value, which is less
than previously estimated for this source. 

We detected gas outflowing at $500-1000$\,km\,s$^{-1}$ as shown by strong, blueshifted absorption lines at different ionization states that are well described with photoionized absorbers. We ascribed the outflow to the photoionized winds that have been observed in several sources (see e.g. \citealt{DiazTrigo2012}). These winds are thought to originate from thermal and radiation pressure near the accretion disk. Our velocity estimates are in agreement with the dynamics of the accretion disks, but our ionization parameters are slightly lower than usual. However, we notice that outflows of intermediate-ionization gas have also been observed.

The Ne/O abundance ratio in the wind is supersolar which would suggest high Ne/O abundances for the donor star if the wind originates from the accretion disk. This would disagree with the hypothesis of the companion to be a brown dwarf unless the matter is expelled from the stellar interior.

{We also detected a system of emission lines in the spectrum of the source after a very powerful burst was exhibited. These lines can be well modeled with collisionally-ionized gas,  originating presumably closer to the compact object, but they require very high Doppler velocities ($v\sim0.1c$). Additional, higher-resolution X-ray (MEG) and far-UV (HST) observations covering the $1-20$\,{\AA} and $1000-2000$\,{\AA} spectral ranges can improve the abundance estimates of the winds and the interpretation of the emitting plasma.}

Our method can be easily tested in several X-ray binaries in order to search for wind signatures and to probe the local circumstellar and the diffuse interstellar media.

\bibliographystyle{aa}
\bibliography{bibliografia} 

\end{document}